\documentclass{article}
\usepackage{geometry}                
\usepackage{graphicx}
\usepackage{stmaryrd}
\usepackage{amssymb}
\usepackage{amsthm, bm}
\usepackage{bbm}
\usepackage{amsmath, cancel, centernot }
\usepackage[mathscr]{eucal}
\usepackage{mathtools}
\usepackage{epstopdf}
\usepackage{mathrsfs}
\usepackage{newtxmath}
\usepackage[markup=underlined]{changes}

\title{Rational Quantum Mechanics: Testing Quantum Theory with Quantum Computers}
\author{Tim Palmer\\ Department of Physics, University of Oxford, UK\\
tim.palmer@physics.ox.ac.uk}
\date{\today}                                          
\makeatletter
\newcommand\be{\@ifstar{\[}{\begin{equation}}}
\newcommand\ee{\@ifstar{\]}{\end{equation}}}
\newcommand\bp{\begin{pmatrix}}
\newcommand\ep{\end{pmatrix}}

\newtheorem*{theorem*}{Theorem}

\makeatother
\begin{document}
\bibliographystyle{plain}
\maketitle
\abstract{Motivated in part by John Wheeler's assertion that the continuum nature of Hilbert Space conceals the `it-from-bit' information-theoretic character of the quantum wavefunction, a theory of quantum physics (Rational Quantum Mechanics - RaQM) is proposed based on a specific discretisation of complex Hilbert Space. The Schr\"{o}dinger equation is not modified in RaQM, even during measurement. However, the bases in which the quantum state is defined must satisfy certain rational-number constraints. These constraints lead to the notion of finite qubit information capacity $N_{\mathrm{max}}$: for any $N > N_{\mathrm{max}}$ qubit state, there is insufficient  information in the $N$ qubits (linearly growing in $N$) to allocate even one bit to each of all $2^{N+1}-2$ continuum degrees of freedom (exponentially growing in $N$) associated with quantum mechanics/theory (QM, where $N_{\mathrm{max}}=\infty$). It is proposed that the discretisation of Hilbert Space in RaQM is due to gravity, hence QM is the (singular) continuum limit of RaQM at $G=0$. On this basis, it is estimated that $N_{\mathrm{max}}$ lies between about 200 and 400 for current qubit technologies, and will never exceed 1,000. Whilst QM and RaQM are experimentally indistinguishable for small numbers of qubits, RaQM predicts that the exponential advantage of quantum algorithms which, like Shor's, require bases with maximal $N$-qubit superposition/entanglement, will have saturated at 1,000 perfect qubits. Hence, insofar as a classical computer will never factor a 2048-bit RSA integer, RaQM predicts that a quantum computer won't either. This predicted breakdown of QM could be testable in less than 5 years. 

\section*{Significance Statement}
\emph{Is there a fundamental reason why quantum computers cannot factor large integers used for encryption today? We introduce a theory of quantum physics based on the notion that the continuum nature of quantum mechanics' state space approximates something inherently discrete, and argue that the reason for such discreteness is gravity.  From this we predict that quantum computers will never break realistic RSA-encrypted messages, for fundamental rather than practical reasons. This predicted breakdown of quantum theory may be falsifiable in a few years. If verified, quantum computers' biggest impact may be in the development of new finite theories which synthesise quantum and gravitational physics, potentially with commercial benefits (albeit for future generations) eclipsing those derived from quantum mechanics alone.}

\section{Introduction}

It has been reported \cite{Gidney} that 2048-bit RSA integers could be factored in under a week by a quantum computer with less than a million noisy qubits. But could a quantum computer with such capability ever be built? Unitary quantum mechanics/theory (QM) does not itself limit the number of qubits that can be coherently entangled in quantum computers. In addition, environmental decoherence can be minimised by ensuring a quantum computer's qubits are sufficiently isolated from their environmental, e.g., by positioning them, as has been suggested, in a deep crater on the dark side of the Moon. 

However, perhaps there are other constraints on what a quantum computer can in principle achieve. For example, gravitationally induced state collapse would certainly limit multi-qubit coherence \cite{Diosi:1989} \cite{Ghirardi} \cite{Penrose:1996}. However, for a million entangled quantum-computing qubits, collapse timescales are longer than the age of the universe and therefore irrelevant as a constraint on near-future quantum computing capability. Another source of error might be the indirect dissipative effects of state collapse \cite{Vischi}. However, no dissipative effects have been observed \cite{Donaldi}. 

Hence, it would appear that limitations on quantum computing performance are technical and not fundamental. However, here we propose a novel type of fundamental constraint that is not directly related to state collapse and involves no dissipative effects: the notion that a qubit has a finite information capacity. To be clear about what is being proposed here, consider a quantum mechanical qubit state 
\begin{equation}
\label{qubit2}
|\psi(\theta, \phi)\rangle= \cos \frac{\theta}{2} |1\rangle + e^{i \phi} \sin\frac{\theta}{2} |-1\rangle
\end{equation}
in the eigenvector basis  $(|1\rangle, |-1\rangle)$ of some hermitian operator, with 2 continuum degrees of freedom represented by $\theta, \phi \in \mathbb R$. Certainly in QM, $\cos \theta$ and $\phi/2\pi$ could be irrational numbers. Indeed, they might be non-computable numbers whose full description would require the specification of unbounded amounts of information. QM does not itself limit the number-theoretic complexity associated with these two degrees of freedom. 

Does this matter? In classical physics, we routinely discretise differential equations to solve them numerically on a digital computer. However, unless these equations are pathological, theorems exist \cite{Strogatz} which assert that for fine-enough discretisation, the numerical solution will be as close as we like to the continuum solution. In this sense, the differential equations define a smooth limit for the discretised models as the discretisation scale goes to zero. However, we cannot assume the continuum limit is smooth if we discretise QM. In particular, one of Lucien Hardy's axioms for QM \cite{Hardy:2004} - the single most important in his estimation - asserts the continuity of QM's state space: complex Hilbert Space. This raises the possibility that QM is a singular limit \cite{Berry} of some fundamentally new theory of quantum physics in which Hilbert Space is discretised \cite{Buniy:2005} \cite{Hsu:2021} \cite{Carroll}, as the discretisation scale goes to zero. Although such a theory may have similar observational and experimental consequences to QM, Hardy's axiom suggests it will have an \emph{extremely different} conceptual underpinning to QM (e.g, as general relativity is to Newtonian gravity). The question we consider here is whether such a theory has testable consequences, different from QM. By way of example, consider a model of fluid turbulence based on the viscous Navier-Stokes equations at very large but finite Reynolds number $Re$, and a model based on the inviscid Euler equations. The two theories generate similar types of turbulent motion. However, in going from Navier-Stokes to Euler, the boundary condition at an aerofoil fundamentally changes, from no slip to no normal motion. This has testable consequences: aircraft cannot fly in an inviscid fluid!

With this in mind, we will suppose that the continuum nature of complex Hilbert Space is merely an idealisation for - and hence an approximation to -  a deeper, inherently discrete theory of quantum physics. Such a possibility was advocated by the eminent physicist John Wheeler, who famously coined the aphorism `It from Bit' (i.e., reality from information), writing \cite{Wheeler}: 
\begin{quote}
The familiar probability function or functional, and wave equation or functional wave equation, of standard quantum theory provide mere continuum idealizations and by reason of this circumstance conceal the information-theoretic source from which they derive.
\end{quote}
As such, we present a theory of quantum physics `Rational Quantum Mechanics - RaQM' based on a discretisation of Hilbert Space, itself based on a discretisation of the Riemann Sphere of (extended) complex numbers. Specifically, in RaQM, permitted qubit bases, i.e., bases where $|\psi\rangle$ in (\ref{qubit2}) is mathematically defined, must be associated with rational values of $\cos \theta$ and $\phi/2\pi$ (more precise restrictions are defined below). Of course, we cannot assume that Wheeler would have agreed with this particular choice of discretisation, however, as discussed below, it does indeed reveal the information-theoretic nature of the qubit state, concealed by QM. As part of this revelation, we define a finite quantity $N_{\mathrm{max}}$ referred to as Qubit Information Capacity (QIC): when $N > N_{\mathrm{max}}$, there is insufficient bitwise information contained in an $N$-qubit system to allocate even one bit of information to each of the $2^{N+1}-2$ Hilbert-space dimensions demanded by QM. In QM itself, $N_{\mathrm{max}}=\infty$, hence QM's unbounded QIC poses no fundamental constraint on experimental quantum physics.

A sufficiently large but finite $N_{\mathrm{max}}$ implies that RaQM is experimentally indistinguishable from QM for systems comprising a small number of qubits. However, due to the exponential growth of Hilbert Space with $N$, any finite $N_{\mathrm{max}}$ could constrain the behaviour of a quantum system with sufficiently many entangled qubits, providing the entanglement is fully spread across the available state space. The purpose of this paper is to propose an experimental test (which might be achievable in as few as 5 years) of the existence of such a constraint. 

But why take RaQM seriously, finiteness issues notwithstanding? As discussed in a companion paper \cite{Palmer:2026a}, RaQM provides a wholly novel way of framing and hence understanding the so-called `weird' properties of quantum physics, such as complementarity, non-commutativity of observables and violation of Bell's inequality. This reframing illustrates the singular nature of QM as a limiting case of RaQM. By abandoning continuity \emph{even slightly}, one can arrive at a theory of quantum physics with radically different mathematical and conceptual properties to QM itself. Not least, by violating the Measurement Independent assumption in Bell's Theorem \cite{HossenfelderPalmer}, not for nominal measurement settings under the full control of the experimenter, but for \emph{exact} measurement settings which were never under their control anyway (consider a gravitational wave passing through an interferometer at the time of measurement), RaQM can be shown to describe a  locally realistic theory of quantum physics.

Information-theoretic considerations notwithstanding, it is proposed that the key \emph{physical} reason for eschewing the state-space continuum (and hence the infinitesimal \cite{tHooft:2015} \cite{Ellis:2018} \cite{Gisin:2021}) is gravity. It is generally accepted that gravitational effects may to a breakdown of the continuum structure of space-time; \emph{a fortiori} we propose such effects lead to a breakdown of the continuum structure of state space. As such, we claim that gravity should play a fundamental role in the very structure of theories of quantum physics, and hence is not some additional field `to be quantised'. Consistent with this, we use existing models of gravitised QM \cite{Penrose:2014} to provide quantitative estimates of the discretisation scale and hence QIC, for different types of qubit. 

The specific framework for discretising Hilbert Space is outlined in Section \ref{discretise}. Based on a constructive discretisation of the Riemann Sphere of extended complex numbers described in the Supporting Information, an explicitly information-theoretic representation of the $N$-qubit state in discretised Hilbert Space is presented in Section \ref{inform}. As discussed in Section \ref{reduce}, state reduction/measurement in discretised Hilbert Space is represented geometrically by a state-space magnification of a Cantor Set, and arithmetically by a chaotic shift map on 2-adic representations of the Cantor Set. By equating the information-theoretic reduction time with the Di\'{o}si-Penrose collapse time \cite{Diosi:1989} \cite{Penrose:1996}, we arrive in Section \ref{estimate} at a simple formula for QIC. $N_{\mathrm{max}}$ for different types of qubit in existence today are estimated and an upper bound of $N_{\mathrm{max}} \approx 1,000$ that will never be exceeded is found. The proposed test of RaQM and hence of the breakdown of QM itself is described in Section \ref{test}. Concluding remarks are made in Section \ref{conclusions}, where we collect together the assumptions made in developing RaQM. 

\section{Discretised Hilbert Space}
\label{discretise}

RaQM invokes a restriction to the Dirac-von Neumann axioms of QM. In particular, a qubit state (\ref{qubit2}) in RaQM, evolved using the Schr\"{o}dinger equation without modification, is only mathematically defined in bases $\{|1\rangle, |-1\rangle\}$ where
\begin{equation}
\label{rat}
\cos^2 \frac{\theta}{2}= \frac{m}{L} \in \mathbb Q ; \ \ \ \ \frac{\phi}{2\pi}= \frac{n}{L} \in \mathbb Q.
\end{equation}
$L \in \mathbb N$ is the single most important variable in this paper (a function of the mass/energy of the qubit as defined in Sections \ref{reduce} and \ref{estimate}), defining the degree of granularity of discretised Hilbert Space; $0 \le m, n \le L$ are whole numbers. Qubits in quantum computers will be found (Section \ref{estimate}) to have $L \ggg 1$. By contrast, the classical limit corresponds to `maximal discretisation' at $L=1$. QM (where $|\psi\rangle$ is mathematically defined in the bases of all hermitian operators) corresponds to the singular limit of RaQM at $L=\infty$. As mentioned in the Introduction, a model with finite $L$ leads to a complete reimagining of the conceptual foundations of quantum physics. For example, as discussed below, Born's Rule is emergent in RaQM and hence not needed as a separate axiom.

By construction, $|\psi\rangle$ is undefined in a basis where the rationality constraint (\ref{rat}) is not satisfied - for example, where $\cos \theta =2\cos^2 (\theta/2) -1$ and/or $\phi/2\pi$ are irrational numbers. Such hypothetical bases are found to be associated ubiquitously with counterfactual measurements which arise when analysing non-classical properties of quantum systems such as complementarity, non-commutativity of observables, and entanglement (e.g., when considering the outcome of a simultaneous counterfactual which-way experiment, when in reality one has performed an interferometric experiment on some specific photon) \cite{Palmer:2026a}.

Equation (\ref{rat}) is straightforwardly generalised for multi-qubit states. In QM, a general $N$-qubit state can be written
\begin{equation} 
\label{alpha}
\alpha_{1} |{1,\ldots,1, 1}\rangle +  \alpha_2 |{1,\ldots, 1, -1}\rangle
 + \alpha_3 |{1, \ldots, -1,1}\rangle
+\ldots + \alpha_{2^N} |{-1,\ldots, -1,-1}\rangle.
\end{equation}
where $\alpha_i \in \mathbb C$. (\ref{alpha}) can be written in an explicitly normalised form; e.g. for $N=3$, as
\begin{equation}
\label{threeraqm}
\begin{split}
\underbrace{\cos\frac{\theta_1}{2} |1\rangle}_1\times 
&\left(
\begin{array}{cc}
\underbrace{\cos \frac{\theta_2}{2}|1\rangle}_2\  
\times\  (\underbrace{\cos\frac{\theta_4}{2}|1\rangle+e^{i \phi_4} \sin\frac{\theta_4}{2}|-1\rangle}_3)+ \\
+\underbrace{e^{i \phi_2}\sin \frac{\theta_2}{2}|-1\rangle}_2
\times (\underbrace{\cos\frac{\theta_5}{2}|1\rangle+e^{i \phi_5} \sin\frac{\theta_5}{2}|-1\rangle}_3) 
\end{array}
\right) +\\
+\underbrace{e^{i \phi_1}\sin\frac{\theta_1}{2} |-1\rangle}_1 \times
&\left(
\begin{array}{cc}
\underbrace{\cos \frac{\theta_3}{2}|1\rangle}_2
\ \times \ (\underbrace{\cos\frac{\theta_6}{2}|1\rangle+e^{i \phi_6} \sin\frac{\theta_6}{2}|-1\rangle}_3)+ \\
+\underbrace{\ e^{i \phi_3} \sin \frac{\theta_3}{2}|-1\rangle}_2
\times (\underbrace{\cos\frac{\theta_7}{2}|1\rangle+e^{i \phi_7} \sin\frac{\theta_7}{2}|-1\rangle}_3)
\end{array}
\right) 
\end{split}
\end{equation}
where
\begin{align}
\label{rats}
\alpha_1&=\cos (\theta_1/2) \cos (\theta_2/2) \cos (\theta_4/2) \nonumber \\
\alpha_2&=\cos(\theta_1/2)\cos(\theta_2/2) \sin (\theta_4/2) e^{i \phi_4}\nonumber \\
\ldots \nonumber \\
\alpha_7&=\sin(\theta_1/2)\sin(\theta_3/2) \sin (\theta_7/2) e^{i (\phi_1+\phi_3+\phi_7)}
\end{align}
The form (\ref{threeraqm}) is represented schematically in Fig \ref{nested}, where the 3 underbraced nested qubits in (\ref{threeraqm}) exhibit 2, 4 and 8 degrees of freedom respectively, totalling 14. Continuing to larger values of $N$, each new qubit adds $2^N$ extra degrees of freedom to the quantum state, yielding $2+4+8+ \ldots +2^N=2^{N+1}-2$ in total ($=2^N$ complex degrees of freedom, less 2 for normalisation and global phase as in (\ref{alpha})). In RaQM, we generalise (\ref{rat}) so that it applies to each of the $\cos^2 \theta_i/2$ and each of the $\phi_i/2\pi$. 
\begin{figure}
\centering
\includegraphics[scale=0.6]{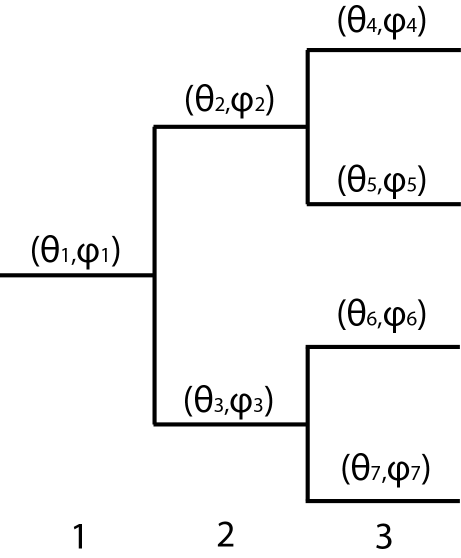}
\caption{\emph{A schematic representation of (\ref{threeraqm}) corresponding to (\ref{alpha}), explicitly normalised, for $N=3$. The self-similar generalisation of this representation to arbitrary $N$ is straightforward, giving $2+4+8+\ldots 2^N=2^{N+1}-2$ degrees of freedom in total. Here the discretisation ansatz (\ref{rat}) is applied to each pair $(\theta_i, \phi_i)$.}}
\label{nested}
\end{figure}

As suggested by Fig \ref{nested}, the form (\ref{threeraqm}) is similar to that for a photon (say) passing through a set of nested beamsplitters. The notion of using multiple beam splitters as a way of testing the author's model of discretised Hilbert Space is described in \cite{Hance:2025}. The estimates below, e.g., of $N_{\mathrm{max}} \approx 1,000$ seem so large that building a one-off laboratory apparatus seems impractical. On the other hand, piggy-backing on the quantum-tech revolution, building such an apparatus may not be inconceivable in the future. In any case, we argue below that quantum computers (no matter what qubit technology is used) provide a testbed for discretised Hilbert Space, potentially in a few years. 

Before continuing, we consider a question whose answer may seem esoteric, but will be central, not only to the mathematical viability of RaQM, but also to the way in which the notion of measurement is described in RaQM. Since rational and irrational numbers can be infinitesimally close, isn't RaQM (where states in bases with irrational squared amplitudes and/or phases are undefined) necessarily extremely fine tuned, and hence not robust to small perturbations? To see that this is \emph{not} the case, first note that irrational real numbers are limit points of Cauchy sequences of rational numbers with specific respect to the Euclidean metric. However, it is well known that Cauchy sequences of rationals can also be defined using the so-called $p$-adic metric \cite{Katok}, with respect to which the irrational reals are simply undefined (and hence infinitely far away). The $p$-adic metric is to fractal geometry as the Euclidean metric is to Euclidean geometry. In particular, the set of 2-adic integers is homeomorphic to the self-similar Cantor Set $\mathscr C$ \cite{Katok}\cite{Robert}. With this in mind, when we come to discuss state reduction as a state-space instability (Section \ref{reduce}), it will be done with respect to a magnification of $\mathscr C$, corresponding to a shift map on the 2-adic numbers which label pieces of $\mathscr C$. All this goes to reinforce the idea that QM is the \emph{singular} continuum limit of RaQM at $L=\infty$. 

\section{Information-theoretic Description of the $N$-Qubit State in RaQM}
\label{inform}

As already discussed, discretising Hilbert Space, even when $L \ggg 1$, leads to a very different theoretical description of the quantum wavefunction than in QM. Specifically, the qubit state (\ref{qubit2}) in a basis satisfying the rationality constraints (\ref{rat}), can be expressed as a finite length-$L$ bit string \cite{Palmer:2020}. The key to this result, as described explicitly in the Supporting Information, is a discretisation of the Riemann Sphere of (extended) complex numbers as permutation/negation operators acting on length-$L$ bit strings. On the discretised Riemann Sphere, $i=\sqrt{-1}$ is not introduced axiomatically as in $\mathbb C$, but constructively, as a permutation/negation operator acting on 2-bit strings $\{a_1, a_2\}$ where $a_i \in \{1,-1\}$. Specifically, $i\{a_1,a_2\}=\{-a_2, a_1\}$ so that $i^2\{a_1, a_2\}=-\{a_1, a_2\}$, where $-\{a_1, a_2\}=\{-a_1, -a_2\}$. The full 3D discretised Riemann Sphere is built up from $i$, using constructive quaternionic/spinorial generalisations of $i$, as described in the Supporting Information. The discretised Riemann Sphere has non-trivial arithmetic properties which underpin RaQM's unique physical interpretation (e.g., quantum physics without randomness or spooky action at a distance; see Section \ref{conclusions} and \cite{Palmer:2026a}).

The discretised Riemann Sphere leads directly to the discretised Bloch Sphere as $L$-bit strings. Specifically, 
\begin{align}
\label{bits}
|\psi(\theta, \phi) \rangle \equiv \{\underbrace{\ 1,\; \ \; 1,\ \; 1,\ \ \dots \ \ 1,\ \; 1,\ \; 1,\ \ -1,-1,-1, \ \ldots \ ,-1,-1,-1}_{\theta, \phi}\}\ \ \ \bmod \xi 
\end{align}
where the bits `1' and '-1' are symbolic labels for measurement outcomes \cite{Schwinger}. Here the fraction of 1s in $|\psi\rangle$ is equal to $\cos^2 (\theta/2)$ and $\phi/2\pi=n/L$ encodes $n$ cyclic permutations of the $L$-bit string. $\xi$ is a specific but unknown permutation operator, representing a specific but unknown relationship of a particular quantum system with respect to the rest of the universe. In RaQM, $\xi$ acts as a hidden variable in RaQM, accounting for what in QM would be described as the inherent randomness of quantum measurement. Just as $|\psi(\theta, \phi)\rangle$ in QM is invariant under a global phase transformation, so $|\psi(\theta, \phi)\rangle$ in (\ref{bits}) is invariant under a generic permutation $\xi$. As such, $|\psi\rangle$ in (\ref{bits}) describes an unordered ensemble of $L$ possible binary measurement outcomes for identically prepared states, with frequencies consistent with Born's rule. 

Importantly, however, for some specific $\xi$, RaQM allows the state of some particular quantum system to be defined (something problematic in QM) as the specific bit string
\begin{align}
\label{bits2}
|\psi(\theta, \phi) \rangle_\xi = \xi \{\underbrace{\ 1,\; \ \; 1,\ \; 1,\ \ \dots \ \ 1,\ \; 1,\ \; 1,\ \ -1,-1,-1, \ \ldots \ ,-1,-1,-1}_{\theta, \phi}\}
\end{align}
which can be written as the difference between two complementary integers in standard base-2 representation (for example, $\{1,-1,-1,1\} \equiv 1001.-0110.$). As will be discussed elsewhere, such a representation allows one to describe single-particle interference in a Mach-Zehnder interferometer - the key point being that the finite bit-string representation of that single quantum particle is the resource that can describe its wave-like properties. 

A system comprising $N$ entangled qubits is represented as $N$ correlated length-$L$ bit strings, where the same permutation $\xi$ applies to each string (consistent with $\xi$ as a global phase in QM). Importantly, $N$ qubits comprise $N \times L$ bits of information. For example, with $N=3$, in some particular rational basis, the 3 length-$L$ bit strings corresponding to (\ref{threeraqm}) can be written as
\be
\label{3bit}
|\psi\rangle \equiv \left\{
\begin{array} {c}
\{\underbrace{\ 1,\; \ \; 1,\ \; 1,\ \ \dots \ \ 1,\ \; 1,\ \; 1,\ \ -1,-1,-1, \ \ldots \ ,-1,-1,-1}_{\theta_1, \phi_1}\}\ \ \ \bmod \xi \\
\{\underbrace{1,1,\; \ldots , 1, \ -1, -1,\ldots,-1}_{\theta_2, \phi_2} \ \underbrace{\ 1,1, \ldots ,\ \ 1,\ -1,-1 \ldots \ ,-1}_{\theta_3, \phi_3}\}\ \ \ \bmod \xi  \\
\{\underbrace{1,\ \ \ldots \ \ ,-1\;}_{\theta_4, \phi_4} \;\underbrace{\ \ 1,\ \ \ldots \ \ ,-1}_{\theta_5, \phi_5}\ \ 
\underbrace{\ 1,\ \ \ldots \ ,-1}_{\theta_6, \phi_6} \ \underbrace{\ 1,\ \ \ldots \ \  ,-1}_{\theta_7, \phi_7}\}\ \ \ \bmod \xi
\end{array}
\right.
\ee
where the variables in the under-braces correspond directly to the continuum degrees of freedom in (\ref{threeraqm}) and Fig \ref{nested}, but now subject to the rationality constraints (\ref{rat}). For example, $\cos^2 \theta_2/2$ denotes the fraction of $1$ bits in the second bit string which correspond to 1 bits in the first bit string, and $\phi_2$ represents a cyclic permutation of these specific bits in the second bit string. As above, these bit strings can be interpreted as an ensemble of measurement outcomes described by the 8 symbolic labels $(1,1,1)$, $(1,1,-1)$, $\ldots$, $(-1,-1,-1)$, i.e., as belonging to the set $\{\pm1,\pm 2,\pm 3,\pm 4\}$. 

As with a single qubit, it is also possible to to interpret a set of bit strings, for a specific $\xi$, as representing the state 
\be
\label{3bitxi}
|\psi\rangle_\xi = \left\{
\begin{array} {c}
\xi \{\underbrace{\ 1,\; \ \; 1,\ \; 1,\ \ \dots \ \ 1,\ \; 1,\ \; 1,\ \ -1,-1,-1, \ \ldots \ ,-1,-1,-1}_{\theta_1, \phi_1}\}  \\
\xi \{\underbrace{1,1,\; \ldots , 1, \ -1, -1,\ldots,-1}_{\theta_2, \phi_2} \ \underbrace{\ 1,1, \ldots ,\ \ 1,\ -1,-1 \ldots \ ,-1}_{\theta_3, \phi_3}\}  \\
\xi \{\underbrace{1,\ \ \ldots \ \ ,-1\;}_{\theta_4, \phi_4} \;\underbrace{\ \ 1,\ \ \ldots \ \ ,-1}_{\theta_5, \phi_5}\ \ 
\underbrace{\ 1,\ \ \ldots \ ,-1}_{\theta_6, \phi_6} \ \underbrace{\ 1,\ \ \ldots \ \  ,-1}_{\theta_7, \phi_7}\}
\end{array}
\right.
\ee
of an individual 3-qubit quantum system. These can be represented, as above, as a sum of complementary integers in base-4 (e.g., $\{2,-3,-1,4\}\equiv 2004. - 0310.$.)

At $L = \infty$ (the QM limit), arbitrarily many degrees of freedom can be encoded in the infinitely long bit strings, and there would be no information-theoretic constraint on the number of qubits that could be entangled. However, if $L$ has some finite value, there will be a finite limit to the number of degrees of freedom that can be encoded in these bit strings. In particular, since the total number of the number of bits in $N$ bit strings equals $N \times L$, and the number of degrees of freedom in an $N$-qubit state in QM is $2^{N+1}-2$, then when 
\be
\label{est}
2^{N+1}-2 > N \times L
\ee
there simply aren't enough bits to allocate even one bit to each QM degree of freedom. By way of illustration, suppose $L=16$. A 5-qubit state in QM has 62 degrees of freedom. Since 5 length-16 bit strings contain 80 bits in total, there are just enough bits in the 5 bit strings to allocated at least 1 bit to each QM degree of freedom. However, a 6-qubit state in QM has 126 degrees of freedom. Since 6 length-16 bit strings contain 96 bits in total, there aren't enough bits to allocate even 1 bit to each degree of freedom. Hence, with $L=16$, $N_{\mathrm{max}}=5$. 

When $L \gg 1$, (\ref{est}) implies
\be
N_{\mathrm{max}} \approx \log_2 L
\ee
In the classical limit $L=1$ there aren't enough bits for even $N_{\mathrm{max}}=1$: the classical limit is not quantum-like at all! Consistent with this, as shown in the Supporting Information, the discretised Riemann Sphere contains no complex structure when $L=1$. 

\section{State Reduction/ Measurement}
\label{reduce}

Although state reduction plays no direct role in defining QIC (and hence on the constraint on quantum computers), state reduction in RaQM is analysed here as it allows us to estimate $L$ numerically (see Section \ref{estimate}). In RaQM, state reduction can be defined in two equivalent ways. First, it describes a steady loss of information in a quantum system as that system becomes entangled with the environment. Secondly, it describes a chaotic state-space instability, separating initially close state-space trajectories until they are classically (see e.g., Fig 30.19 of \cite{Penrose:2004}). To illustrate these processes in RaQM, we return to the discussion at the end of Section \ref{discretise} and consider the Cantor Set $\mathscr C$ (see Fig \ref{measurement}). The iterates of $\mathscr C$ are themselves labelled by $L$, where coarser iterations correspond to smaller values of $L$. This is consistent with the representation of pieces of $\mathscr C$ by 2-adic numbers - the coarser the iteration of $\mathscr C$, the fewer the number of bits needed to represent the corresponding 2-adic integers. We use an initial value $L=4$ in Fig \ref{measurement} purely for illustrative convenience - the generalisation to much larger values of $L$ is entirely straightforward. The two pieces of $\mathscr C$ at $L=1$ are to be considered attractors to which the qubit state is attracted. The fractal structure of $\mathscr C$ implies that the corresponding basins of attraction are profoundly intertwined \cite{Palmer:1995}.

\begin{figure}
\centering
\includegraphics[scale=0.6]{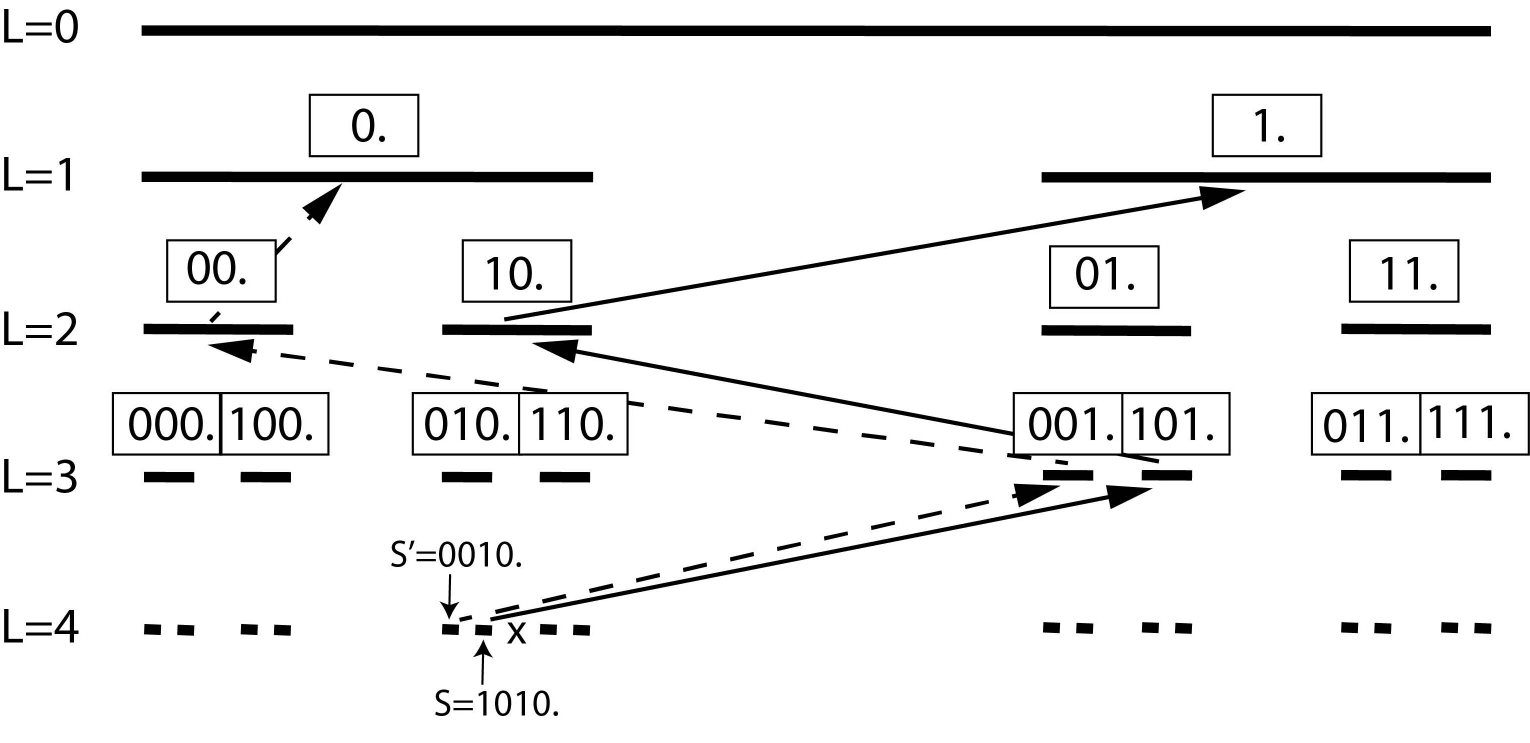}
\caption{\emph{State reduction as both loss of information and as a chaotic state-space instability, describable geometrically as a magnification of a Cantor Set $\mathscr C$, or arithmetically as a shift map on the 2-adic representation of a rational number. Here, for illustrative simplicity, we consider a qubit state $S$ (relative to a basis satisfying (\ref{rat})) initially as a piece of $\mathscr C$ at the $L=4$th iteration. The iterate at $L=3$ shows a magnification of the left half of the iterate at $L=4$ containing the state $S$, and the iterate at $L=2$ is a magnification of the right half of the iterate at $L=3$, and so on. This process is represented arithmetically by the shift map (\ref{shift}). Whilst $S=1010.$ reduces to $1.$, the neighbouring (2-adically close) state $S'=0010.$ reduces to $0.$, illustrating the nature of the instability. The hypothetical state represented by `$X$' to the immediate right of $S$, but not lying on $\mathscr C$, is completely undefined. $X$ might correspond to the qubit state in a counterfactual basis where squared amplitudes or complex phases (in degrees) are irrational numbers. Such bases typically occur when considering non-classical properties of quantum systems \cite{Palmer:2026a}.}}
\label{measurement}
\end{figure}

Here a qubit state $S$ is represented by a small segment of $\mathscr C$ at the $L=4$th iterate, or equivalently by a 2-adic integer with 4 bits (Fig \ref{measurement}). State reduction to the classical limit $L=1$ occurs in 3 timesteps $t_*$. During the first timestep, from $L=4$ to $L=3$, the left half of the Cantor Set (containing $S$) is magnified back to $\mathscr C$'s original size. During the second timestep, the right half of the $L=3$rd iterate of $\mathscr C$ (containing $S$) is again magnified back to $\mathscr C$'s original size. Finally, during the third timestep, the left half of the $L=2$nd iterate of $\mathscr C$ is again magnified back to $\mathscr C$'s original size. $S$ is now `reduced' to the right-hand piece of $\mathscr C$ at $L=1$. Shown is the corresponding evolution of the state $S'$ initially neighbouring $S$ at $L=4$, to the left-hand piece of $\mathscr C$ at $L=1$, illustrating geometrically the nature of the instability. 

Equivalently, such state reduction is given by a simple chaotic shift map \cite{WoodcockSmart} on the 2-adic representation of $S$, each step corresponding to a division by 2. In particular, under the 3 timesteps, $S$ is mapped 
\be
\label{shift}
1010. \mapsto 101. \mapsto 10.\mapsto 1. \ ,
\ee
consistent with Fig \ref{measurement} and highlighting measurement as a loss of information (to the environment). With respect to the 2-adic metric, $S'=0010.$ is 2-adically close to $S=1010.$ and hence neighbours $S$ on $\mathscr S$ (Fig \ref{measurement}). Under the 3 timesteps $t_*$, $S'$ is mapped to 
\be
0010. \mapsto 001. \mapsto 00.\mapsto 0. \ ,
\ee
This is consistent with $S$ and $S'$ `reducing' to the two classically distinguishable attractors at the $L=1$ iterate. The evolution to $L=1$ is associated with a decrease over $3t_*$ from the initial value $L=4$. For a general $L$, the collapse time $\tau_{\mathrm{*}}$ to $L=1$ is given by 
\be
\label{collapse}
\tau_{\mathrm{*}} = \left (L -1 \right ) t_*
\ee
i.e., $L-1$ individual magnifications/shift maps, terminating at $L=1$. 

But what corresponds to the timestep $t_*$? This question can be answered by assuming that the physical process associated with Hilbert Space discretisation is gravity. It is already widely assumed that the space-time continuum will break down at the Planck Scale. Perhaps, \emph{a fortiori} a synthesis of quantum and gravitational physics will lead to a breakdown in the Hilbert Space continuum. Since QM has been such experimentally well-tested theory, it is clear that any such breakdown of the Hilbert Space continuum must occur on very fine scales (in the sense that discretisation is undetectable for small numbers of qubits). This is consistent with gravity being so weak. However, as stressed above, if the continuum limit is singular, there will surely be some observable consequences of Hilbert Space discretisation, no matter how is fine the discretisation. 

With this assumption, we draw on the Di\'{o}si-Penrose \cite{Diosi:1989} \cite{Penrose:1996} model of quantum state reduction with collapse timescale
\begin{equation}
\label{DP}
\tau_{\mathrm{DP}}=\frac{\hbar}{E_G}
\end{equation}
Here, $E_G$ is the Di\'{o}si-Penrose gravitational self-energy associated with two instances of a mass in QM superposition. As discussed e.g., in \cite{Penrose:2014}, $E_G$ can be viewed as the energy needed to move one instance of the mass away from the other by a distance $b$ taking account of the gravitational field of the masses. It can be noted that when $M$ equals the Planck mass, $\tau_{\mathrm{DP}}=t_P$, the Planck time. 

We now let $\tau_{\mathrm{*}}= \tau_{\mathrm{DP}}$ and put $t_*=t_P$. Then $L$ is described by the simple formula
\begin{equation}
\label{limit}
L  =\lceil \frac{E_{P}}{E_G}\rceil
\end{equation}
where $E_{P}$ is the Planck energy ($\approx 10^9$J) and $\lceil x \rceil$ denotes the ceiling function mapping the real number $x$ to the nearest integer greater than $x$. The identification $t_*=t_P$ is consistent with the Di\'{o}si-Penrose model, since if $E_G$ were very slightly less than $E_P$, so that $L=2$, reduction to the classical limit would occur in one timestep $t_*=t_P$ according to the Di\'{o}si-Penrose model. From (\ref{limit}), we see that QM corresponds to the singular (unphysical) limit $E_G=0$, i.e. $G=0$. 

It is important to note that in RaQM, state collapse is not described by modifying the Schr\"{o}dinger equation. In particular, this state-reduction process does not itself imply any local space-time energy dissipation, a feature of some collapse models (e.g., \cite{Pearle:1976} \cite{Ghirardi}). Hence RaQM is wholly consistent with a failure to detect the local effects of energy dissipation during measurement \cite{Donaldi}. 

As $L$ decreases to 1, we expect the quantum system to become more and more entangled with its environment. As discussed in Section \ref{inform}, an $N$-qubit system in RaQM is described by $N$ partially correlated bit strings, or equivalently as 1 string comprising $2^N$ labels.  Hence, the state of the fully entangled system in the classical limit $L=1$ is describable by a single one of these $2^N$ labels, i.e. as a single integer $p$ (typically $\gg 0$) lying between 0 and $2^N$. Since these are now single integers, and not strings of integers, In describing how different systems described by different integers $p_1$, $p_2$ etc interact when $L=1$, the arithmetic would necessarily be Euclidean rather than $p$-adic. In this sense, the classical limit corresponds not only to $L=1$, but to a transition from $p$-adic to Euclidean arithmetic. We will pursue this conclusion elsewhere. 

The state reduction process described here plays no direct role in the quantum computing constraint described in this paper. Rather, we `merely' invoke state reduction in order to estimate QIC numerically. To see the irrelevance of state reduction as a direct physical constraint, consider our estimate (Section \ref{estimate}) $L =10^{70}$ for a quantum dot qubit. After 1 billion years, $L$ will have decreased by $\approx 3 \times 10^{60}$ which is a negligible fraction of $10^{70}$. By contrast, if the qubit had sufficient mass that $L \approx 10^{40}$, still $\gg 1$, then collapse to $L=1$ would take less than 1ms. 

\section{Estimating $L$ and $N_{\mathrm{max}}$.}
\label{estimate}

For a system of mass $M$ with characteristic size $R$, in QM superposition over a distance $b$, we make use of formulae for $E_G$ already available in the literature. In particular \cite{Howl}:
\begin{align}
\label{EG}
E_G&= \frac{6GM^2}{5R} \left( \frac{5}{3} \beta^2 - \frac{5}{4} \beta^3 + \frac{1}{6} \beta^5 \right ) \ \ \mathrm{if} \ \ \ 0 \le \beta \le 1 \nonumber \\
&=\frac{6GM^2}{5R} \left( 1- \frac{5}{12\beta} \right )  \ \ \mathrm{if} \ \ \ \beta \ge 1 \nonumber \\
\end{align}
where $\beta= b/(2R)$. 

1) Consider a typical quantum dot associated with an electron of mass $10^{-30}$kg in QM superposition over $b=5$nm. A question immediately rises as to whether we take $R$ to be the size $R_e$ of the electron, $10^{-15}$m, or the size $R_p$ of the electron's QM probability cloud, say $10^{-9}$m. We will use this ambiguity to provide an estimate of uncertainty in $E_G$. In both cases $\beta \ge 1$ and, for order-of-magnitude estimates we use the approximate expression
\be
\label{EGG}
E_G \approx \frac{GM^2}{R} \approx 10^{-55}\ J 
\ee
when $R=R_e$ and $E_G \approx 10^{-61}$J when $R=R_p$. Using (\ref{limit}), then $L \approx  10^{64} \approx 2^{212}$ when $R=R_e$ and $L \approx  10^{70} \approx 2^{232}$ when $R=R_p$. In both cases, we have $N_{\mathrm{max}} \approx 200$, showing that the calculation is not especially sensitive to the choice of $R$. 

2) With $t_*=t_P$, the state-reduction model for discretised Hilbert Space necessarily incorporates the speed of light $c$. We will take advantage of this to estimate $L$ for photonic qubits in a quantum computer, in superposition over a distance given by some fraction $\beta$ of the photon wavelength $\lambda$. Although the Di\'{o}si-Penrose model was developed within a Newtonian gravitational framework, we assume it will continue to hold in a post-Newtonian framework where gravity couples to energy. For a photon, we let $M$ in (\ref{EG}) denote the relativistic mass $h/c \lambda $ in the computer's rest frame. Consider an infrared photon of wavelength $\lambda =10^3$nm, as used in a photonic quantum computer. With $\beta = 10^{-3}$ say, (\ref{EG}) becomes
\be
\label{EGGG}
 E_G \approx \beta^2 \frac{G h^2}{c^2 \lambda^3}\approx 10^{-82}\ J
 \ee
so that $L \approx 10^{91} \approx 2^{302}$. Hence, $N_{\mathrm{max}} \approx 300$, somewhat bigger than for the quantum dot example. To be consistent with special relativity, both $L$ and $\tau_*$ are frame dependent.  

3) Consider an ion in an ion-trap where the qubit basis states are based on hyperfine atomic line splitting with an energy difference $\Delta E \approx10^{-9}$eV. Writing (\ref{EGGG}) with $\beta \approx 1$, then 
\be
\label{EGGGG}
E_G \approx \frac{G \Delta E^3}{h c^5}\approx 10^{-100}\ J
\ee
and hence $L \approx 10^{109} \approx 2^{400}$. Hence $N_{\mathrm{max}} \approx 400$, somewhat higher than the estimates above. 

4) Undoubtedly, technological developments in qubit precision will increase $N_{\mathrm{max}}$ further. However, is there an upper limit to $N_{\mathrm{max}}$ that technology, no matter how precise, cannot exceed? Consider the energy $10^{-32}$eV of a photon with the lowest possible frequency that can be accommodated within the age of the universe (taken as $10^{11}$ years), giving a wavelength $\lambda = 10^{25}$m. Now consider two such ultra-low frequency photons in QM superposition over a distance of one Planck length $10^{-35}$m so that $\beta^2 \approx 10^{-120}$. Using (\ref{EGGG}), $E_G \approx 10^{-289}$J and hence 
$L \approx 10^{298}$, giving $N_{\mathrm{max}}\approx 1,000$. 

These estimates are based on single qubit mass/energy values. In the limit where the qubits in a quantum computer are spatially well separated, then the relevant $E_G$ for the composite system will be equal to that for the individual qubits. On the other hand, if we were to consider the $N$ qubits as a single `conglomerated' mass, this would slightly reduce the estimates of $N_{\mathrm{max}}$ but not by much. For example, if $M \mapsto 300 M$ in (\ref{EGG}) then $E_G \approx 2^{17} E_G$, and the estimate of $N_{\mathrm{max}}$ only decreases by 17. 

This limit does not change the more a quantum system is shielded from its non-gravitational environment; positioning a quantum computer in a deep crater on the dark side of the moon will not help to overcome the constraints of a finite $N_{\mathrm{max}}$. By the Principle of Equivalence, a quantum computer will always be gravitationally coupled to the rest of the universe.  


\section{An Experimental Test of RaQM}
\label{test}

From the discussion above, and the values obtained for $L$, RaQM will be experimentally indistinguishable from QM for small numbers of entangled qubits, and indeed for large numbers $N>N_{\mathrm{max}}$ qubits where superposition/entanglement is restricted to a low-dimensional sub-space of the full $2^{N+1}-1$ Hilbert Space dimension. On the other hand RaQM will be experimentally distinguishable from QM for $N>N_{\mathrm{max}}$ for quantum systems where superposition/entanglement is maximally distributed across the full $2^{N+1}-2$ dimensions of Hilbert Space. 

The Quantum Fourier Transform is such an algorithm \cite{Tyson:2003}. Hence a candidate for testing RaQM (and hence the breakdown of QM) is Shor's algorithm. For example, the quantum circuit described in \cite{Willsch}, implementing Shor's algorithm, needs $N$ qubits to factor an $N-1$-bit semi-prime. Here $N=2,049$ exceeds even our largest conceivable estimate of $N_{\mathrm{max}}$. Hence, insofar as it is impossible to factor a 2048-bit integer (within the age of the universe) using a classical computer, RaQM predicts it will be impossible to factor such an integer with a quantum computer. Indeed, according to estimates described in Section \ref{estimate}, a quantum computer based on current qubit technology will already have ceased having an advantage over a classical computer in factoring a 1,024-bit RSA integer. 

In practice, the performance of quantum computers is limited by qubit imperfections and environmental noise. Error-correction algorithms can correct for such factors, utilising additional qubits. Hence, RaQM does not predict that the performance of noisy qubits will saturate at 1,000 qubits. However, suppose, based on QM, it will be possible to factor a 2,048-bit RSA integer in a quantum computer using some large number $N_{2,048}>2,048$ of noisy qubits (say a million \cite{Gidney}), then we predict, based on RaQM, that the performance of the computer in factoring such an integer will be no better than a classical computer as the number of qubits approaches $N_{2,048}$. We therefore have a test of RaQM that could be falsified in a few years if quantum technology roadmaps are accurate. 

Perhaps it might be argued that macroscopic quantisation contradicts predictions of the breakdown of QM when $N>N_{\mathrm{max}}$. For example, a superconducting fluid comprising Avogadro's number $N_e$ of electrons is very well described by QM (e.g., using the Bardeen, Cooper Schrieffer BCS model) where $N_e \gg N_{\mathrm{max}}$. On the other hand, whilst such a system has the potential for each of its component parts to be describable by individual wave functions so that the system is a product state of a very large number of individual subsets, in practice these subsets lock into a single collective wave function, having, for example, the same phase. In this way, one can talk about \emph{the} phase of the system's wave function independent of the individual electrons. In the BCS model, therefore, the system evolves in a very small subset of the available Hilbert Space. That is to say, unlike the Quantum Fourier Transform, the BCS model does not involve bases in which the system's state is maximally spread across the available Hilbert Space.  A classical analogy of the point being made here might be the nonlinear synchronisation of a large number $N_o$ of oscillators. The fact that the oscillators are synchronised means that the phase portrait of the system can be described in a very much smaller dimensional space than that of the Cartesian product of all $N_o$ individual oscillators. As such, macroscopic quantisation does not contradict RaQM's predictions. 

QM is not predicted to break down `suddenly' at $N = N_{\mathrm{max}}$ qubits. For $N< N_{\mathrm{max}}$, then $N L / (2^{N+1}-2)$ bits can be allocated to each Hilbert Space dimension. For example, with $N_{\mathrm{max}} \gg 1$ and $N=N_{\mathrm{max}}-1$, then $2^1$ qubits can be allocated to each Hilbert State dimension. When $N$ is 5 qubits short of $N_{\mathrm{max}}$, then $2^5=32$ bits can be allocated to each Hilbert Space dimension, and so on. Hence, if the computer can be completely shielded from environmental noise, the output from the computer (in factoring integers with Shor's algorithm) will appear to be subject to an increasingly strong type of discretised `Shot' noise as $N$ approaches $N_{\mathrm{max}}$. 

\section{Conclusions and Discussion}
\label{conclusions}

A novel theory, Rational Quantum Mechanics (RaQM), of quantum physics has been ourlined. In RaQM, the complex Hilbert state of an entangled $N$-qubit system, evolved using the unmodified Schr\"{o}dinger equation, is only mathematically defined in bases where squared amplitudes and complex phases (in degrees) are rational numbers of the form $m/L$ where $L \in \mathbb N$ and $m$ is a whole number with $m \le L$. Using a constructive discretisation of the Riemann Sphere of complex numbers, the complex Hilbert state in such a basis can be described as $N$ correlated length-$L$ bit strings, where $L$ is a fundamental variable of RaQM describing the degree of granularity of Hilbert Space. QM is itself the singular continuum limit of RaQM at $L=\infty$. 

There are a number of compelling reasons for positing such a theory, only some of which are discussed in this paper. Firstly, as discussed, it allows the information theoretic nature of the quantum system to be described explicitly \cite{Wheeler}. Secondly, as also discussed, it provides an explicit description of state reduction to the classical limit $L=1$ in terms of a chaotic instability which can be described both geometrically and arithmetically. Thirdly, not discussed here but see the companion paper \cite{Palmer:2026a}, see also below, discrete Hilbert Space provides novel interpretations of inherently quantum properties, such as complementarity, non-commutativity of observables and entanglement (and hence EPR/Bell nonlocality). Finally, as discussed, discretisation provides a novel route for incorporating the effects of gravity in quantum physics, not as some field to be quantised by independently derived quantisation rules, but as an inherent ingredient of such quantisation rules. 

Based on RaQM, we predict a breakdown of QM that may be experimentally falsifiable in a few years using quantum computers. The assumptions made in making this prediction are summarised here:
\begin{itemize}
\item RaQM assumes the discretisation ansatz (\ref{rat}) for a single qubit, thus introducing the finite discretisation parameter $L$, also named Qubit Information Capacity (QIC). 
\item  RaQM assumes a straightforward generalisation for multi-qubit states, implying that the QIC of $N$ entangled qubits equals $N \times L$. Hence, for given $L$, QIC grows linearly with $N$. Since in QM the Hilbert space dimension of $N$ entangled qubits equals $2^{N+1}-2$ and hence grows exponentially with $N$, there must exist a maximum value $N_{\mathrm{max}} \approx \log_2 L$ of $N$, above which it is not possible to allocate even one bit of information to each QM degree of freedom. At this point, a quantum algorithm such as Shor's, which utilises Hilbert Space maximally, can no longer continue to have an advantage over a corresponding classical algorithm.
\item We estimate QIC by studying state reduction. It is assumed that state reduction is associated with a chaotic state-space instability whereby initially nearby states diverge exponentially until they have separated sufficiently to be classically distinguishable. This process corresponds to a decrease in $L$ by a value of 1 per timestep $t_*$ until the classical limit at $L=1$ is reached. Physically, we can suppose that this reduction in information is lost to the environment as the quantum system decoheres.  
\item It is assumed that gravity sets the discretisation scale $L$, so that the state reduction timescale becomes $\tau_*=(L-1)t_P$, where $t_P$ is the Planck time. Consistent with this, $\tau_*= \tau_{\mathrm{DP}}$, the Di\'{o}si-Penrose collapse timescale, so that $L = \lceil E_P/E_G \rceil$, where $E_P$ is the Planck energy, $E_G$ is the Di\'{o}si-Penrose gravitational self-energy difference, and $\lceil \ldots \rceil$ is the ceiling function of number theory. 
\item Existing formulae for $\tau_{\mathrm{DP}}$ are used to estimate $L$ for a quantum dot qubit in a quantum computer, giving  $N_{\mathrm{max}} \approx 200$.
\item It is assumed that the Diosi-Penrose formula can be extended to describe photonic qubits by replacing $M$ with the relativistic mass-energy $h/c\lambda$ in the computer's rest frame. This leads to a value $N_{\mathrm{max}} \approx 300$ for a photonic qubit, and $N_{\mathrm{max}} \approx 400$ for an ion trap qubit based on hyperfine line splitting. 
\item Considering a photon with frequency given by the inverse age of the universe, put into phase superposition over one Planck length, we show that $N_{\mathrm{max}}$ can never exceed $1,000$ and hence 2048-bit RSA integers will never be factored.  
\end{itemize}

The notion that there may be an information-theoretic limitation to the capability of a quantum computer is not entirely new (e.g., \cite{tHooft:2015b}). In particular Davies \cite{Davies:2007} has argued, using Lloyd's \cite{LLoyd:2002} estimate $I_{\mathrm{universe}} \approx 10^{122}$ of bits of information (in bosons, fermions and gravitational degrees of freedom) encoded in the universe, that the Hilbert Space dimension of a quantum state which entangles more than $N_{\mathrm{Davies}}= \log_2 I_{\mathrm{universe}}$ qubits is larger than $I_{\mathrm{universe}}$. Davies suggests that this may signal a fundamental physical limit to quantum computation. 

Whether one is convinced by a cosmological argument of this type or not, the value $N_{\mathrm{Davies}} \approx 400$ is remarkably similar to the ones deduced here, made without explicit reference to the rest of the universe. Is this a coincidence? The author thinks not. The original motivation for developing a model of quantum physics based on discretised Hilbert Space \cite{Palmer:2020} was as a way to understand the violation of Bell's inequality. For example, the arithmetic properties of the discretisation of the Riemann Sphere (see Supporting Information) imply that the Measurement Independence postulate \cite{HossenfelderPalmer} in Bell's Theorem is violated, not for nominal measurement settings that are always under the full control of experimenters, but for exact measurement settings that were never under their control anyway. Hence, as discussed in \cite{Palmer:2026a}, the violation of Bell's inequality need not imply abandonment of local realism. However, it must imply some notion of holism. Indeed, one can take inspiration from the development of general relativity, motivated by Mach's holistic but not nonlocal principle that inertia `here' is due to mass `there'. The author's Invariant Set Postulate \cite{Palmer:2009a}, that the universe be considered a chaotic dynamical system evolving precisely on a dynamically invariant measure-zero fractal, is also an example of a holistic but not EPR/Bell nonlocal concept - and is consistent with the $p$-adic state representations discussed in this paper. In this regard, the close numerical agreement between $N_{\mathrm{max}}$ and $N_{\mathrm{Davies}}$ may well be evidence of the holistic nature of the laws of physics at their most fundamental, as discussed above. This surely must be important when attempting to synthesise quantum and gravitational physics. 

Much of the development of quantum computers has been motivated by their potential commercial applications. If the proposed test does indeed support RaQM, it will stymie the development of such applications. On the other hand, if quantum computers can aid the development of an improved understanding of the fundamental laws of physics, not least in helping to synthesise quantum and gravitational physics, then currently unforeseeable commercial opportunities may arise for future generations. 

\section*{Acknowledgements}

My thanks to Simon Benjamin, Stephen Blundell, Paul Davies, Chris Heunen, Sabine Hossenfelder, Richard Howl, Roger Penrose, Felix Tennie and the PNAS reviewers for extremely helpful comments and inputs. My thanks also to Brendan Palmer for assistance with the graphics. The research described in this paper was supported by the Leverhulme Trust Grant Number EM-2023-060. 

\bibliography{mybibliography}

\section*{Supporting Information}\label{secA1}
\subsection*{Discretising the Riemann Sphere}
\label{Riemann}

The continuum field $\mathbb C$, with its axiomatically defined element $i=\sqrt{-1}$, plays an essential role in QM (either explicitly or implicitly as in the de Broglie-Bohm interpretation). Central to RaQM is a discretisation of the Riemann Sphere $\mathbb C \cup \{\infty\}$ of (extended) complex numbers $z(\theta, \phi)=\cot (\theta/2)e^{i \phi}$ where $\theta$ denotes co-latitude (or zenith angle) and $\phi$ denotes longitude (or azimuth). The discretisation is defined by
\begin{equation}
\label{ratL}
\cos^2 \frac{\theta}{2}= \frac{m}{L} \in \mathbb Q ; \ \ \ \ \frac{\phi}{2\pi}= \frac{n}{L} \in \mathbb Q.
\end{equation}
where $L \in \mathbb N$ and $0 \le m, n \le L$ are whole numbers. Note that if $\theta$ satisfies (\ref{ratL}) then since $\cos \theta = 2 \cos^2 (\theta/2)-1$, necessarily $\cos \theta \in \mathbb Q$. $L$ is an especially important variable, defining the degree of granularity of discretised Hilbert Space: QM is the (singular) limit of RaQM at $L=\infty$, whilst the classical limit of RaQM occurs at $L=1$. The key property of this discretisation is that the $i$ (with property $i^2=-1$) does not have to be introduced by axiom. Rather, as discussed below, $i$ and the corresponding quaternions are defined constructively from representations of complex numbers as permutation/negation operators acting on bit strings. With this definition of $i$, the entire set $\{z(\theta, \phi)\}$ of complex numbers on the discretised Riemann Sphere is obtained constructively by an inductive argument, for all $L \in \mathbb N$. 

\subsubsection*{Discretised Complex Numbers and Quaternions}

When $L=2^0=1$, (\ref{ratL}) implies $\theta \in  \{0, \pi\}$, $\phi =0$, i.e., the discretisation only admits 2 points at the north and south poles. We represent the corresponding set of `complex numbers' in terms of the identity and negation operators
\be
\label{L1}
z(0, 0) \{1\}=\{1\};\ \ z(\pi, 0)\{1\}=\{-1\}=-\{1\}
\ee
With $L=2^0$ (the classical limit of RaQM), the discretisation is too coarse to admit complex structure at all. Complex numbers play no essential role in classical physics. 

When $L=2^1$, then $\theta \in \{0, \pi/2, \pi\}$, $\phi \in \{0, \pi\}$ according to (\ref{ratL}). $L=2$ is large enough to admit complex structure, represented by the permutation/negation operator (PNO) $i$ defined as
\be
\label{i}
 i\{a_1,a_2\}=\{-a_2, a_1\}
 \ee
where $a_i \in \{1, -1\}$, so that $i^2\{a_1,a_2\}=\{-a_1, -a_2\} \equiv -\{a_1, a_2\}$. Generalising (\ref{L1}), the set of complex numbers on the $L=2$ discretised Riemann Sphere are associated with the 4 PNOs $\{i^0, i, i^2, i^3\}$. Specifically, 
\begin{align}
&z(0, 0)\{1,1\}=i^0\{1,1\}=\{1,1\}; \nonumber \\
&z(\frac{\pi}{2}, 0)\{1,1\}=i\{1,1\}=\{-1,1\}; \nonumber \\
&z(\pi, 0)\{1,1\}=z(\pi,\pi)\{1,1\}=i^2 \{1,1\}=\{-1,-1\}; \nonumber \\
&z(\frac{\pi}{2}, \pi)\{1,1\}=i^3\{1,1\}=\{1,-1\}
\end{align}
as shown in Fig \ref{Riemann1}a.  

\begin{figure}
\centering
\includegraphics[scale=0.65]{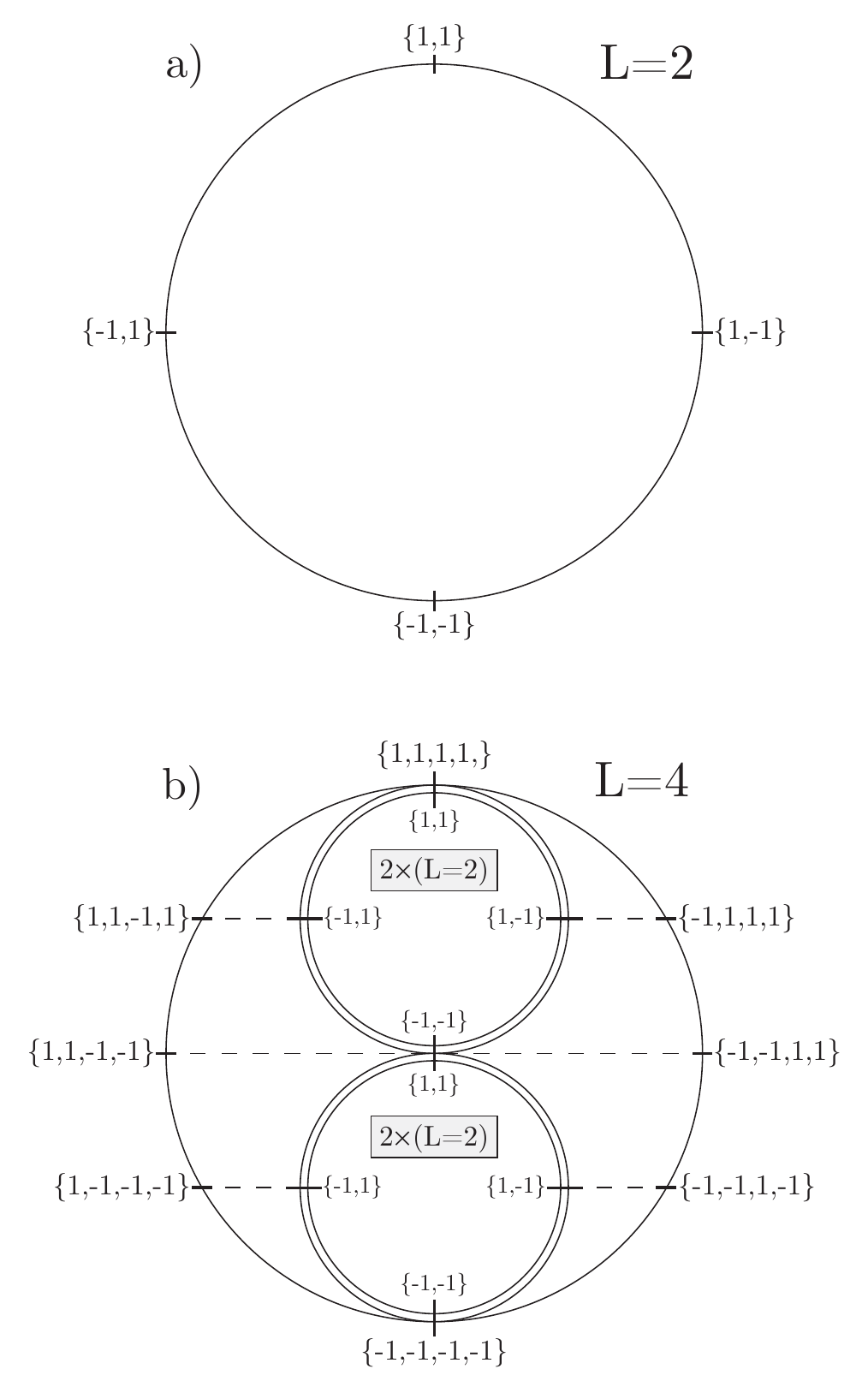}
\caption{\emph{a) The discretised Riemann Sphere at $L=2$, represented as four 2-bit strings on a complete great circle at $\phi=0,\pi$ through the poles. b) A schematic illustration of how to construct the discretised Riemann Sphere for $L=4$, again on the great circle through $\phi=0,\pi$, from two $L=2$ great circles as in a). See the text for details. This also illustrates the construction of an $L=2^M$ discretised Riemann Sphere on the great circle at $\phi=0, \pi$, from two $L/2$-bit discretised great circles. }}
\label{Riemann1}
\end{figure}

When $L=2^2=4$ then $\theta \in \{0, \pi/3, \pi/2, 2\pi/3, \pi\}$, $\phi \in \{0, \pi/2, \pi, 3\pi/2\}$ according to (\ref{ratL}). The discretised sphere now comprises 8 points on each of two complete great circles (at $\phi=0, \pi$ and $\phi=\pi/2, 3\pi/2$) passing through the north and south poles - 14 points in total. $L=4$ is now large enough to admit quaternionic structure, as shown below. The complex numbers at these points are represented as PNOs acting on 4-bit strings. We start by considering the points on the great circle $\phi=0,\pi$, writing $\{1,1,1,1\}$ as two concatenated 2-bit strings $\{1,1\} |\{1,1\}$. We define PNOs based on $\{i^0, i, i^2, i^3\}$ acting individually on each pair of 2-bit strings, with a construction motivated by the theory of spinors. Specifically, we first apply $i$ twice to the second 2-bit string keeping the first fixed. We then apply $-i$ twice to the first 2-bit string keeping the second fixed; then again apply $i$ twice to the second bit string keeping the first fixed; then again apply $-i$ twice to the first bit string keeping the second fixed. In total, from Fig \ref{Riemann1}a, the first 2-bit string has been rotated by $4\pi$ relative to the second (like Dirac scissors relative to the back of the chair). Specifically, 
\begin{align}
z(0, 0)\{1,1,1,1\}=&\; i^0\{1,1\}|i^0 \{1,1\}=\{1,1,1,1\}; \nonumber \\
z(\frac{\pi}{3}, 0)\{1,1,1,1\}=&\;  i^0\{1,1\} | i^1 \{1,1\}=\{1,1,-1,1\};\nonumber \\
z(\frac{\pi}{2}, 0)\{1,1,1,1\}=&\; i^0 \{1,1\} | i^2 \{1,1\}=\{1,1,-1,-1\};\nonumber\\
z(\frac{2\pi}{3}, 0)\{1,1,1,1\}=&\; i^3 \{1,1\} | i^2 \{1,1\}=\{1,-1,-1,-1\}; \nonumber \\
z(\pi,0)\{1,1,1,1\}=z(\pi,\pi)\{1,1,1,1\}=&\;  i^2 \{1,1\} | i^2 \{1,1\}=\{-1,-1,-1,-1\}; \nonumber \\
z(\frac{2\pi}{3}, \pi)\{1,1,1,1\}=&\;  i^2 \{1,1\} | i^3 \{1,1\}=\{-1,-1,1,-1\};\nonumber \\
z(\frac{\pi}{2},\pi)\{1,1,1,1\}=&\;  i^2 \{1,1\} |\ i^0 \{1,1\}=\{-1,-1,1,1\};\nonumber \\
z(\frac{\pi}{3},\pi)\{1,1,1,1\}=&\; i^1 \{1,1\} | i^0 \{1,1\}=\{-1,1,1,1\}
\end{align}
This construction is illustrated schematically in Fig \ref{Riemann1}b. Here the 4-bit strings on the $L=4$ great circle comprise two 2-bit strings from the pair of smaller $L=2$ circles, where one rotates in steps of $\pi/2$ by $4 \pi$ relative to the other. 

The complex numbers on the $\phi=\pi/2, 3\pi/2$ great circle are PNOs obtained by a cyclic permutation of bit strings on the $\phi =0, \pi$ great circle, so that
\begin{align}
z(0, \frac{\pi}{2})\{1,1,1,1\}=&\; \{1,1,1,1\}; \nonumber \\
z(\frac{\pi}{3}, \frac{\pi}{2})\{1,1,1,1\}=&\; \{1,-1,1,1\}; \nonumber \\
z(\frac{\pi}{2}, \frac{\pi}{2})\{1,1,1,1\}=&\; \{1,-1,-1,1\};\nonumber\\
z(\frac{2\pi}{3}, \frac{\pi}{2})\{1,1,1,1\}=&\; \{-1,-1,-1,1\}; \nonumber \\
z(\pi,\frac{\pi}{2})\{1,1,1,1\}=z(\pi,\frac{3\pi}{2})=&\; \{-1,-1,-1,-1\}; \nonumber \\
z(\frac{2\pi}{3}, \frac{3\pi}{2})\{1,1,1,1\}=&\; \{-1,1,-1,-1\}; \nonumber \\
z(\frac{\pi}{2},\frac{3\pi}{2})\{1,1,1,1\}=&\; \{-1,1,1,-1\}; \nonumber \\
z(\frac{\pi}{3},\frac{3\pi}{2})\{1,1,1,1\}=&\; \{1,1,1,-1\}.
\end{align}
That this cyclic permutation is consistent with quaternionic structure can be seen by defining the operators
\begin{align}
I \{a_1, a_2, a_3, a_4\} =  \{a_3, a_4, -a_1, -a_2\} \nonumber \\
J \{a_1, a_2, a_3, a_4\} =  \{a_2, -a_1, -a_4, a_3\} \nonumber \\
K \{a_1, a_2, a_3, a_4\} =  \{-a_4, a_3, -a_2, a_1\} 
\end{align}
which satisfy 
\begin{equation}
I^2=J^2=K^2=- 1; \ \ \ I \times J=K.
\end{equation}
In particular, $I\{1,1,1,1\}=\{1,1,-1,-1\}=z(\pi/2,0)\{1,1,1,1\}$ and $J \{1,1,1,1\}=\{1,-1,-1,1\}=z(\pi/2, \pi/2)\{1,1,1,  1\}$, consistent with $I$ and $J$ rotating the 4-bit string $\{1,1,1,1\}$ at the north pole to the 4-bit strings on the equator at $\phi=0$ and $\phi=\pi/2$, respectively. Consistent with this, $K$ maps one these equatorial points to the other: 
\be
KI\{1,1,1,1\}=K\{1,1,-1,-1\}=\{1,-1,-1,1\}=J\{1,1,1,1\}.
\ee
The 14 4-bit strings on the 3D $L=4$ discretised Riemann sphere are shown in Fig \ref{Riemann2}.
\begin{figure}
\centering
\includegraphics[scale=0.8]{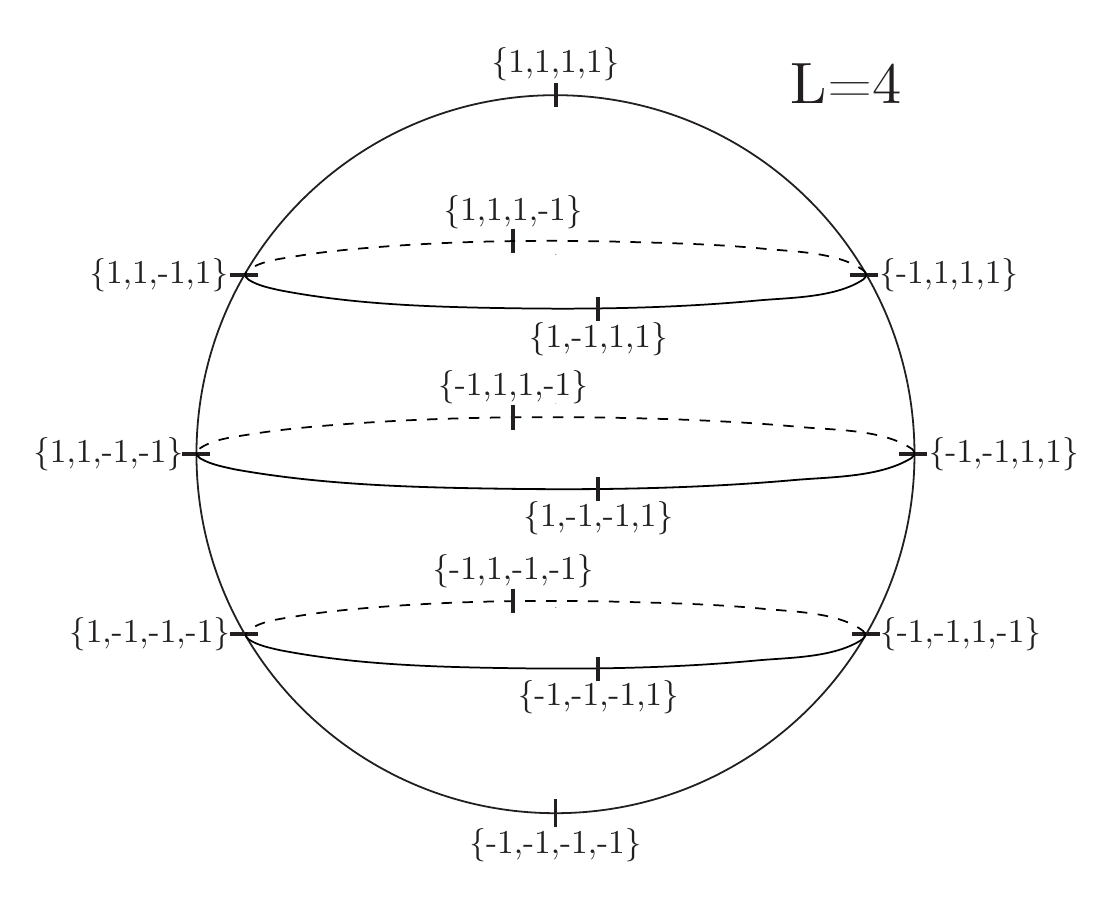}
\caption{\emph{The full 3D discretised Riemann Sphere for $L=4$ - where, consistent with quaternionic structure (see text), rotation about the polar axes corresponding to a cyclic permutation of the 4-bit strings.}}
\label{Riemann2}
\end{figure}
The inductive step is to write the discretised PNOs for some $L=2^M$, acting on the $L$-bit string $\{1,1,\ldots, 1\}$ as PNOs acting on 2 concatenated $L/2$-bit strings $\{1,1,\ldots 1\}|\{1,1,\ldots, 1\}$. The PNOs on the $\phi=0, \pi$ great circle are defined using the spinorial construction illustrated in Fig \ref{Riemann2}b, generalised so that the two smaller circles represent the two $L/2$ discretised great circles at $\phi=0, \pi$. We then rotate the $\phi=0, \pi$ great circle around the polar axis by cyclically permuting the $L$-bit string as in 
\be
\label{zeta}
\zeta \{a_1, a_2, \ldots, a_L\} \mapsto \{a_2, a_3, \ldots, a_L, a_1\}
\ee
Each application of $\zeta$ corresponds to a rotation by $2\pi/L$ radians. 

This completes the $L$-discretisation of the Riemann Sphere when $L$ is a power of 2. For discretisations where $L$ is not a power of 2, one can simply interpolate between constructions where $L$ is a power of 2. For example, with $L=3$, one can simply remove the last bit from the $L=4$ construction on $\phi=0$, yielding
\begin{align}
&z(0, 0)\{1,1,1\}=\{1,1,1\}; \nonumber \\
&z(\theta_*, 0)\{1,1,1\}=\{1,1,-1,\};\nonumber \\
&z(\pi-\theta_*, 0)\{1,1,1\}=\{1,-1,-1\}; \nonumber \\
&z(\pi,0)\{1,1,1\}=\{-1,-1,-1\}
\end{align}
where $\cos \theta_*=1/3$. For the longitudes $\phi=2\pi/3, 4\pi/3$, one simply performs two cyclic permutations to the 3-bit strings. 

Note that, for all $L$, the proportion of 1s in the bit-string representing $z(\theta, \phi)$ is equal to $\cos^2 \theta/2$. Hence the proportion of $-1$s equals $\sin^2 \theta/2$, and the relative frequency of $1$s to $-1$s is equal to $\cot^2 \theta/2$. This is the basis for the claim that there is no need to postulate Born's Rule in RaQM - it is redundant in the Dirac-von Neumann axioms.  

\subsubsection*{Arithmetic Properties of Complex Numbers on the Discretised Riemann Sphere}
\label{Arithmetic}

Central to the physical interpretation of RaQM are the arithmetic properties of the complex-number PNOs on the $L$-bit discretisation of the Riemann Sphere. In particular, suppose $z_1$ and $z_2$ are complex numbers belonging to the $L$-discretised sphere. The key question we consider in this section is whether $z_1\pm z_2$ belongs to the discretised sphere. We show that, typically, but non-trivially, they do not. 

On the Argand plane, $z_1$, $z_2$ and $z_2-z_1$ correspond to three sides of a triangle with vertices at the origin, at $z_1$ and at $z_2$. We will assume the triangle is non-degenerate, i.e. the angle $\phi$ between $z_1$ and $z_2$ is non-zero. On the Riemann Sphere the corresponding triangle is a spherical triangle with vertices which we label $O$, 1 and 2 (where $O$ corresponds to the south pole). 

Now if $z_1=\cot (\theta_1/2)e^{i \phi_1}$ and $z_2=\cot (\theta_2/2)e^{i \phi_2}$, both satisfy the rationality conditions (\ref{ratL}) then $\cos \theta_1 \in \mathbb Q$, $\cos \theta_2 \in \mathbb Q$ and $\phi = (\phi_2-\phi_1)/2\pi \in \mathbb Q$. If we write $z_2-z_1 = \cot (\theta_3/2)e^{i \phi_3}$, then for $z_2-z_1$ to lie on the discretised Riemann Sphere, we require that $\cos \theta_3 \in \mathbb Q$. 

The cosine rule applied to the triangle $\triangle O12$ can be written
\be
\label{sp}
\cos \theta_{3}= \cos \theta_{1} \cos \theta_{2} + \sin \theta_{1} \sin \theta_{2} \cos \phi
\ee
If $\cos \theta_3 \in \mathbb Q$, then, since the first term on the RHS of (\ref{sp}) is rational, the second term on the RHS must also be rational. Squaring, this implies that 
\be
(1-\cos^2 \theta_{1})(1- \cos^2 \theta_{2}) \cos^2 \phi \in \mathbb Q
\ee
and hence $\cos^2 \phi \in \mathbb Q$ and therefore $\cos 2\phi \in \mathbb Q$. However, we now have a contradiction since we know that $\phi/2\pi \in \mathbb Q$. According to Niven's Theorem \cite{Niven} \cite{Jahnel:2005}, the only values $0 \le \phi < 2\pi$ where $\cos 2\phi$ and $2\phi/2\pi$ are simultaneously rational are:
\be
 2\phi =0, \frac{\pi}{3},  \frac{\pi}{2},  \frac{2\pi}{3},  \pi, \frac{4\pi}{3},  \frac{3\pi}{2},  \frac{5\pi}{3}.
 \ee
Hence, unless $2\phi$ is a multiple of $60^\circ$ \emph{exactly},  $\cos 2\phi$ and $2\phi/2\pi$ cannot be simultaneously rational. Hence, modulo these exceptions, if $z_1$ and $z_2$ satisfy (\ref{rat}), $z_2-z_1$ cannot satisfy (\ref{ratL}). Of course, the same conclusion can be drawn if we add $z_2$ to $z_1$. In the argument above we merely replace $\phi$ with $\pi-\phi$. 

We refer to this result as the `Impossible Triangle Corollary' (to Niven's Theorem). Niven's Theorem is central to RaQM and plays a vital role in the interpretation of complementarity, non-commuting observables and to the violation of Bell inequalities in RaQM \cite{Palmer:2026a}. In these interpretations the mathematically impossible bases (one with irrational squared amplitudes and/or phases) correspond to physically impossible simultaneous counterfactual measurements. 

\subsection*{Uncertainty Principle}

\begin{figure}
\centering
\includegraphics[scale=0.4]{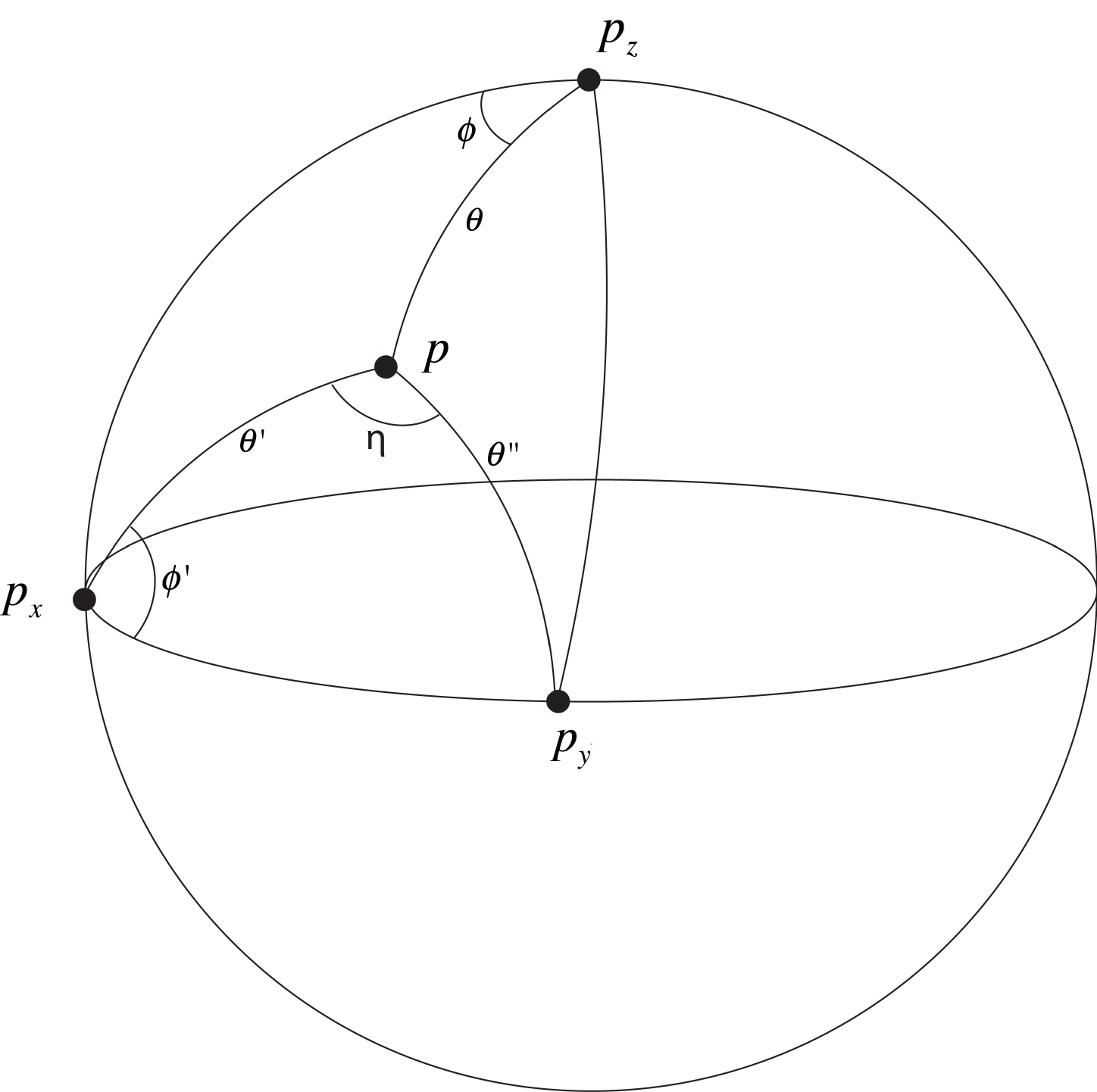}
\caption{\emph{In RaQM, the Uncertainty Principle $\Delta S_x \Delta S_y \ge \frac{\hbar}{2} \overline S_z$  for a single spin qubit arises from the trigonometry of spherical triangles and the correspondence between points on the discretised Bloch sphere with coordinates satisfying the rationality and bit strings.}}
\label{uncertainty}
\end{figure}
In RaQM, the Uncertainty Principle for a single qubit arises from a trigonometric inequality on the sphere, together with Niven's Theorem. We then discuss the Uncertainty Principle for multiple qubits. 

Consider a point $p$ on the unit sphere (Fig \ref{uncertainty}) whose colatitude with respect to the three orthogonal poles $p_x$, $p_y$ and $p_z$ is $\theta$, $\theta'$ and $\theta''$ respectively. The internal angles $\phi'$ and $\eta$ are shown on the figure. By the sine rule for spherical triangle $\bigtriangleup pp_xp_y$
\begin{equation}
\frac{\sin \theta''}{\sin \phi'} = \frac{\sin \pi/2}{\sin \eta}=\frac{1}{\sin \eta} 
\end{equation}
Hence
\begin{equation}
\label{unc1}
|\sin \theta''| \ge |\sin \phi'|
\end{equation}
By the cosine rule for spherical triangle $\bigtriangleup pp_xp_z$, 
\begin{equation}
\label{unc2}
\cos \theta = \sin \theta' \sin \phi'
\end{equation}
From (\ref{unc1}) 
\begin{equation}
|\sin \theta'| |\sin \theta''| \ge |\sin \theta'| |\sin \phi'| 
\end{equation}
and using (\ref{unc2})
\begin{equation}
\label{unc3}
|\sin \theta'| |\sin \theta''| \ge |\cos \theta| 
\end{equation}
As discussed, a bit string at colatitude $\theta$ has a mean value $\mu_{\theta}=\cos\theta$, and standard deviation $\sigma_\theta = \sin \theta$. With this in mind, consider three discretised Bloch spheres, with the north poles oriented at $p_x$, $p_y$ and $p_z$ respectively. With $\cos \theta = \mu_{\theta}$, $\sin \theta = \sigma_{\theta}$ (the mean and standard deviation of the bit string) then from (\ref{unc3}),
\begin{equation}
\label{unc4}
\sigma_{\theta'}\; \sigma_{\theta''} \ge |\mu_{\theta}| 
\end{equation}
If instead of $\pm 1$, the bit strings have dimensional values $\pm \hbar/2$ in order that they correspond to physical spin, then (\ref{unc4}) becomes the familiar uncertainty principle for spin qubits
\begin{equation}
\label{unc5}
\Delta S_x\; \Delta S_y \ge \frac{\hbar}{2}\; \overline S_z
\end{equation}

The rationality constraints play a key role here. If, for example, $\cos \theta \in \mathbb Q$ and $\phi \in \mathbb Q$, by Niven's Theorem applied to $\bigtriangleup pp_xp_z$, $\cos \theta'$ cannot be rational. Hence it is impossible to know simultaneously, the spin values of a particle with respect to the two directions $p_x$ and $p_z$ (similarly for any two other pairs of directions). 

Let us now derive the Uncertainty Principle for conjugate observables defined on multiple qubits (like position and momentum). First we square (\ref{unc4}) and average over $M$ points $p$ on the discretised Bloch Sphere. Then, from (\ref{unc3}), 
\be
\overline{\sigma^2_{\theta'}\; \sigma^2_{\theta''}} \ge \overline{|\mu_{\theta}|^2}
\ee
Now 
\begin{align}
\overline{\sigma^2_{\theta'}} \ \overline{\sigma^2_{\theta''}} =&\frac{1}{M} (\sigma^2_{\theta'_1}+\sigma^2_{\theta'_2} + \ldots + \sigma^2_{\theta'_M}) 
 \times \frac{1}{M} (\sigma^2_{\theta''_1}+\sigma^2_{\theta''_2} + \ldots + \sigma^2_{\theta''_M}) \nonumber \\
&> \frac{1}{M} ((\sigma^2_{\theta'_1}\sigma^2_{\theta''_1}+\sigma^2_{\theta'_2}\sigma^2_{\theta''_2} + \ldots + \sigma^2_{\theta'_M}\sigma^2_{\theta''_M}) 
= \overline{\sigma^2_{\theta'} \sigma^2_{\theta''}}
\end{align}
Hence, if the $M$ points are uniformly distributed with respect to $\cos \theta$ so that $\overline{|\cos \theta|}=1/2$, then
\be
\sqrt{\overline{\sigma^2_{\theta'}}} \ \sqrt{\overline{\sigma^2_{\theta''}}} \ge \frac{1}{2}
\ee
Multiplying by $\hbar$ gives the standard position/momentum form
\be
\Delta x \Delta p \ge \frac{1}{2} \hbar
\ee
 of the Uncertainty Principle. 
 
\end{document}